\documentclass[a4paper,byrevtex,showpacs,nofootinbib,dvips]{revtex4}
\usepackage{dcolumn}
\usepackage{amsmath}
\usepackage{bm}

\begin{document}

%\draft
\title{Three theorems of quantum mechanics and their classical counterparts}

\author{Claude \surname{Semay}}
\email[E-mail: ]{claude.semay@umons.ac.be}
\affiliation{Service de Physique Nucl\'{e}aire et Subnucl\'{e}aire,
UMONS Research Institute for Complex Systems,
Universit\'{e} de Mons,
Place du Parc 20, 7000 Mons, Belgium}

\date{\today}
\begin{abstract}
The Hellmann-Feynman, virial and comparison theorems are three fundamental theorems of quantum mechanics. For the first two, counterparts exist for classical mechanics with relativistic or nonrelativistic kinetic energy. It is shown here that these three theorems are valid for classical mechanics with a nonstandard kinetic energy. This brings some information about the connections between the quantum and classical worlds. Constraints about the functional form of the kinetic energy are also discussed.
\end{abstract}

% 03.65.Ca Quantum mechanics: Formalism 
% 45.20.-d Formalisms in classical mechanics


\maketitle

Let us consider a quantum Hamiltonian of the form 
\begin{equation}
\label{HTV}
H=T(\bm p)+V(\bm r),
\end{equation}
where $V$ is the potential energy depending only on the space variable $\bm r$ and $T$ the kinetic energy depending only on the conjugate momentum $\bm p$. The structure of $T$ cannot be completely arbitrary. Some constraints are given in \cite{Semay2016} for one-dimensional operators. But it is easy to generalize for three dimensions. We can expect that the kinetic energy is a positive quantity which is an increasing function of only the modulus of the momentum: $T(\bm p)=K(p^2)$ with $p=|\bm p|$. Moreover, some degree of differentiability is desirable for the function $K$.

If the eigensolutions are searched for in the $| \bm r \rangle$ representation, the action of $T$ can be computed with the Fourier transform of the wave-function. This is the basis of the Fourier-Grid Hamiltonian method \cite{Marston1989,Semay2000} and a version of the Lagrange-mesh technique \cite{Lacroix2011}. So, it is perfectly relevant to work with a general form $T$ which can differ from the nonrelativistic or relativistic usual ones, and which can be useful for effective models. 

The Hellmann-Feynman theorem \cite{Hellmann1935,Feynman1939} states that, if the Hamiltonian $H(\lambda)$ depends on a parameter $\lambda$, and that $|\lambda\rangle$ is a normalized eigenstate with the energy $E(\lambda)$, then
\begin{equation}
\label{HFtheor}
\frac{d}{d \lambda} E(\lambda) = \left\langle \lambda \left|\frac{\partial H(\lambda)}{\partial \lambda}\right|\lambda\right\rangle.
\end{equation}
A demonstration of a generalized version of this theorem can be found in \cite{Lichtenberg1989}. 

The general virial theorem \cite{Lucha1990} for Hamiltonians of type (\ref{HTV}) states that 
\begin{equation}
\label{virial}
\left\langle \bm p \cdot \frac{\partial T}{\partial \bm p} \right\rangle = \left\langle \bm r \cdot \frac{\partial V}{\partial \bm r} \right\rangle,
\end{equation}
where the mean values are taken with an eigenstate of $H$, and $\partial/\partial \bm a$ is the gradient with respect to the components of $\bm a$. A simple demonstration can be obtained using the Hellmann-Feynman theorem \cite{Ipekoglu2016}.

Let us consider two ordered Hamiltonians $H_1$ and $H_2$, that is to say $\langle \phi| H_1 |\phi \rangle \leq \langle \phi| H_2 |\phi \rangle$ for any state $|\phi \rangle$. 
The comparison theorem states that each corresponding pair of eigenvalues is ordered $E^{(1)}_{\{\alpha\}} \leq E^{(2)}_{\{\alpha\}}$ (${\{\alpha\}}$ represents a set of quantum numbers) \cite{Reed1972}. A simple demonstration can be obtained using again the Hellmann-Feynman theorem \cite{Semay2011}. The existence of an order is easy to check for Hamiltonians of type (\ref{HTV}) \cite{Semay2011}.

The virial and the Hellmann-Feynman theorems are also valid for classical mechanics, even if it is less well known for the latter case \cite{McKinley1971,Perkinsa2015}. The demonstrations are made for relativistic or nonrelativistic kinetic energy operators. Before showing that the three quantum theorems described above are actually applicable to classical mechanics with a general form of $T$, let us study the relevance of such a mechanics. From the classical point of view, the equations of motion for Hamiltonian~(\ref{HTV}) are given by the Hamilton's equations
\begin{align}
\label{rd}
\dot{\bm r} &= \frac{\partial H}{\partial \bm p} = \frac{\partial T}{\partial \bm p}, \\
\label{pd}
\dot{\bm p} &= -\frac{\partial H}{\partial \bm r} = -\frac{\partial V}{\partial \bm r} = \bm F,
\end{align}
where the dot designates the time derivative $d/dt$, and $\bm F$ is the usual force. The energy is constant since 
\begin{equation}
\label{dhdtz}
\frac{d H}{d t}= \frac{\partial H}{\partial \bm p}\cdot \dot{\bm p}+ \frac{\partial H}{\partial \bm r}\cdot \dot{\bm r} = 0,
\end{equation}
thanks to (\ref{rd})-(\ref{pd}). 

Since $T(\bm p)=K(p^2)$, 
\begin{equation}
\label{rdp}
\dot{\bm r} = \frac{\partial K(p^2)}{\partial \bm p}=2\, K'(p^2)\,\bm p.
\end{equation}
In order that $\dot{\bm r}$ and $\bm p$ be aligned with the same direction, $K'(x)$ must be strictly positive (at least for positive values of $x$). Consequently, $K(x)$ is invertible for $x\ge 0$. Eq.~(\ref{rdp}) implies that $\dot{\bm r}^2 = 4\, p^2\, K'(p^2)^2$. If the inverse of the function $4\, x\, K'(x)^2$ exists and is called $Q(x)$, then
\begin{equation}
\label{pvrd}
\bm p = \frac{\dot{\bm r}}{2\,K'(Q(\dot{\bm r}^2))}.
\end{equation}
This equation allows to write (\ref{pd}) as a second order differential equation for $\bm r$, which can be solved by usual techniques. Since $T$ can be written as a function of the variable $\dot{\bm r}$, it is easy to show that (\ref{dhdtz}) gives the usual Euler-Lagrange equations for the Lagrangian $L=T-V$.
The angular momentum $\bm J$ is always defined by 
\begin{equation}
\label{J}
\bm J = \bm r \times \bm p.
\end{equation}
Its derivative with respect to time is still given by the torque
\begin{equation}
\label{Jd}
\dot{\bm J} = \bm r \times \dot{\bm p} = \bm r \times \bm F,
\end{equation}
since $\dot{\bm r} \times \bm p= \bm 0$ by (\ref{pvrd}). 

In order to establish the classical counterparts of the theorems above, it is necessary to find the equivalent of the quantum mean value operation. Following the existence or not of a period for the evolution of the system, two different averaging procedures can be naturally defined. For a closed orbit, a period $\tau$ can be easily determined. So 
\begin{equation}
\label{mvclass1}
\left\langle \cdot \right\rangle_{\textrm{cl}} = \frac{1}{\tau} \int_{t_0}^{t_0+\tau} \cdot\, dt.
\end{equation}
In other situations, a usual definition is
\begin{equation}
\label{mvclass2}
\left\langle \cdot \right\rangle_{\textrm{cl}} = \lim_{\Delta\to\infty}\frac{1}{\Delta} \int_{t_0}^{t_0+\Delta} \cdot\, dt.
\end{equation}
For the virial theorem, one can start from
\begin{equation}
\label{vircl}
\frac{d}{dt}(\bm r\cdot \bm p)= \dot{\bm r}\cdot \bm p + \bm r\cdot \dot{\bm p} = \bm p \cdot \frac{\partial T}{\partial \bm p} - \bm r \cdot \frac{\partial V}{\partial \bm r},
\end{equation}
thanks to (\ref{rd})-(\ref{pd}). For a periodic system of period $\tau$
\begin{equation}
\label{virclperiod}
\left\langle \frac{d}{dt}(\bm r\cdot \bm p) \right\rangle_{\textrm{cl}} = \frac{\left. \bm r\cdot \bm p\, \right|_{t_0}^{t_0+\tau}}{\tau}=0.
\end{equation}
For a non-periodic system 
\begin{equation}
\label{virclnoperiod}
\left\langle \frac{d}{dt}(\bm r\cdot \bm p) \right\rangle_{\textrm{cl}} = \lim_{\Delta\to\infty}\frac{\left. \bm r\cdot \bm p\, \right|_{t_0}^{t_0+\Delta}}{\Delta}=0,
\end{equation}
if the system is bound. So, in both cases, (\ref{vircl}) gives 
\begin{equation}
\label{virclfin}
\left\langle \bm p \cdot \frac{\partial T}{\partial \bm p} \right\rangle_{\textrm{cl}} = \left\langle \bm r \cdot \frac{\partial V}{\partial \bm r} \right\rangle_{\textrm{cl}},
\end{equation}
which is the classical counterpart of (\ref{virial}), that is to say the classical virial theorem.

In the quantum versions of the two other theorems, eigenvalues of different Hamiltonians are  compared by keeping constant the quantum numbers of the corresponding eigenstates. For classical motions, it seems thus natural to keep constant the following action $I$, which is the quantified quantity in the WKBJ approximation \cite{Semay2016},
\begin{equation}
\label{I}
I =\frac{1}{2\pi}\int_C \bm p\cdot d\bm r = \frac{1}{2\pi}\int_C p\, ds,
\end{equation}
where $C$ is a possible trajectory for the particle: a closed curve for a cyclic orbit or a very long trajectory in other cases. Thanks to (\ref{rdp}), $\bm p\cdot d\bm r = p\, \hat{\bm p}\cdot d\bm r = p\, ds$, where $ds$ is a path element. Let us consider the variation of a parameter $\lambda$, among the others kept fixed. Since the Hamiltonians considered depend actually on $p$ and not on $\bm p$, $H=H(\lambda,p,\bm r)$. Consequently $p=p(\lambda,H,\bm r)$ and $I=I(\lambda,H)$. So, the differential $dI$ is written
\begin{equation}
\label{dI}
dI = \frac{\partial I}{\partial H}\, dH  + \frac{\partial I}{\partial \lambda}\, d\lambda .
\end{equation}
The definition chosen for the variation of the energy with a parameter $\lambda$ is 
\begin{equation}
\label{dhdl}
\frac{dE}{d\lambda} = \left.\frac{dH}{d\lambda}\right|_{I\ \textrm{constant}} = - \frac{\partial I/\partial \lambda}{\partial I/\partial H}.
\end{equation}
The denominator and the numerator of this fraction are given by
\begin{equation}
\label{2partial}
\frac{\partial I}{\partial H} = \frac{1}{2\pi}\int_C \frac{\partial p}{\partial H} \, ds \quad \textrm{and} \quad \frac{\partial I}{\partial \lambda} = \frac{1}{2\pi}\int_C \frac{\partial p}{\partial \lambda} \, ds ,
\end{equation}
since the integral is over the whole trajectory. In order to link these integrals with the equations of motion, let us write 
\begin{align}
\label{dH}
dH & = \frac{\partial H}{\partial \lambda}\, d\lambda + \frac{\partial H}{\partial p}\, dp + \frac{\partial H}{\partial \bm r}\cdot d\bm r , \\
\label{dp}
dp & = \frac{\partial p}{\partial \lambda}\, d\lambda + \frac{\partial p}{\partial H}\, dH + \frac{\partial p}{\partial \bm r}\cdot d\bm r .
\end{align}
The comparison of (\ref{dH}) and (\ref{dp}), gives
\begin{equation}
\label{2partialp}
\frac{\partial p}{\partial H} = \frac{1}{\partial H/\partial p}
 \quad \textrm{and} \quad 
\frac{\partial p}{\partial \lambda} = -\frac{\partial H/\partial \lambda}{\partial H/\partial p}.
\end{equation}
Since 
\begin{equation}
\label{partialHp}
\dot{\bm r} = \frac{\partial H}{\partial \bm p} = \frac{\partial H}{\partial p}\,\frac{\partial p}{\partial \bm p}=\frac{\partial H}{\partial p}\,\hat{\bm p},
\end{equation}
it follows that
\begin{equation}
\label{dsdt}
\frac{\partial H}{\partial p} = \hat{\bm p}\cdot \dot{\bm r}=\frac{ds}{dt},
\end{equation} 
and
\begin{equation}
\label{partialIH}
\frac{\partial I}{\partial H} = \frac{1}{2\pi}\int_C  \frac{dt}{ds}\, ds =\frac{\Delta t}{2\pi},
\end{equation} 
where $\Delta t$ is the time taken to travel the trajectory. A similar calculation for $\partial I/\partial \lambda$ gives
\begin{equation}
\label{partialIlambda}
\frac{\partial I}{\partial \lambda} = -\frac{1}{2\pi}\int_C \frac{\partial H}{\partial \lambda} \,\frac{dt}{ds}\, ds = -\frac{1}{2\pi}\int_{t_0}^{t_0+\Delta t} \frac{\partial H}{\partial \lambda} \, dt
\end{equation} 
These results with (\ref{dhdl}) yield the classical Hellmann-Feynman theorem
\begin{equation}
\label{HF}
\frac{dE}{d\lambda} = \left\langle  \frac{\partial H}{\partial \lambda} \right\rangle_{\textrm{cl}},
\end{equation}
in which the integral (\ref{mvclass1}) or (\ref{mvclass2}) is chosen following the characteristics of the trajectory. This demonstration is directly inspired from \cite{McKinley1971,Perkinsa2015}. 

Let us consider two Hamiltonians $H_1=T_1+V_1$ and $H_2=T_2+V_2$, such that $T_2(\bm p) \ge T_1(\bm p)\ \forall\ \bm p$ and $V_2(\bm r) \ge V_1(\bm r)\ \forall\ \bm r$. A new Hamiltonian $H(\mu)$ is defined by $H(\mu)=(1-\mu) H_1+\mu\, H_2$ with $0 \le \mu \le 1$. The energy $E= E(\mu)$ for this Hamiltonian obeys the classical Hellmann-Feynman theorem 
\begin{equation}
\label{Hmu}
\frac{dE}{d\mu} = \left\langle  H_2 - H_1 \right\rangle_{\textrm{cl}}
= \left\langle  T_2 - T_1 \right\rangle_{\textrm{cl}} + \left\langle  V_2 - V_1 \right\rangle_{\textrm{cl}}.
\end{equation}
The quantity $dE/d\mu$ is positive since each integrand is positive everywhere on the trajectory. With $H_1 = H(0)$ and $H_2 =H(1)$, the corresponding energies $E_1$ and $E_2$ are ordered 
\begin{equation}
\label{H12}
E_1 \le E_2,
\end{equation}
provided the value of the action $I$ given by (\ref{I}) is the same for these two solutions. This can be called the classical comparison theorem. The demonstration  above relies on the procedure used in \cite{Semay2011}. 

Though the quantum world is quite different from the classical world, some connections exists, the most famous being certainly the Ehrenfest theorem. It is also possible to define a classical probability distribution which is the result of an appropriate
averaging procedure on a quantum probability distribution \cite{Semay2016}. Another way is discussed here, showing that some quantum theorems have a well-defined classical counterparts for Hamiltonians with a quite general structure.

\end{document}